 \documentclass[onecollarge,natbib]{svjour2}
 \bibpunct{[}{]}{;}{n}{}{,} 
 \smartqed  
 \usepackage{graphicx}
 \journalname{Few-Body Systems}

\usepackage{amssymb}
\usepackage[intlimits]{amsmath}
\usepackage{amsfonts}
\usepackage{dsfont}
\usepackage{slashed}
\usepackage{units}
\usepackage{wrapfig}

      \newcommand{\conjg}[1]{\ensuremath{\hspace{1pt}\overline{\hspace{-1pt}#1\hspace{-1pt}}}\hspace{1pt}}

\usepackage[usenames]{color}
\usepackage{colortbl}

\definecolor{webgreen}{rgb}{0,0.75,0}
\definecolor{webred}{rgb}{0.75,0,0}
\definecolor{webblue}{rgb}{0,0,0.75}
\definecolor{darkblue}{rgb}{0,0,0.6}
\definecolor{darkgreen}{rgb}{0,0.5,0.5}
\definecolor{darkpurple}{rgb}{0.5,0,0.5}
\definecolor{darkorange}{rgb}{1,0.5,0}
\definecolor{darkgrey}{rgb}{0.4,0.4,0.4}
\definecolor{lgray}{rgb}{0.95,0.95,0.95}
\definecolor{lgreen}{rgb}{0.95,1.00,0.90}
\definecolor{lred}{rgb}{1.00,0.90,0.80}
\definecolor{lblue}{rgb}{0.2,0.35,1.00}
\definecolor{shadecolor}{rgb}{1.00,0.92,0.82}

\usepackage[colorlinks=true,linkcolor=lblue,citecolor=lblue,urlcolor=lblue]{hyperref}

\begin{document}

 \title{Towards a microscopic understanding of nucleon polarizabilities\thanks{This work is supported by the German Science Foundation DFG under
         project number DFG TR-16.}
 }


 \author{Gernot Eichmann}


 \institute{G. Eichmann \at
               Institut f\"{u}r Theoretische Physik, Justus-Liebig-Universit\"at Giessen \\
               Heinrich-Buff-Ring 16, 35392 Giessen, Germany \\
               Tel.: +49 641 9933342\\
               Fax: +49 641 9933309\\
               \email{gernot.eichmann@theo.physik.uni-giessen.de}   }

 \date{Received: date / Accepted: date}

 \maketitle

 \begin{abstract}
       We outline a microscopic framework to calculate nucleon Compton scattering from the level of quarks and gluons
       within the covariant Faddeev approach. We explain the connection with hadronic expansions of the Compton scattering amplitude and discuss
       the obstacles in maintaining electromagnetic gauge invariance. Finally we give preliminary results for the nucleon polarizabilities.

 \keywords{Nucleon \and Compton scattering \and Faddeev equations \and Dyson-Schwinger approach}

 \end{abstract}

 \section{Introduction} \label{intro}

       There is much ongoing interest in the precision determination of the nucleon's polarizabilities; see~\cite{Hagelstein:2015egb} for a recent review.
       The electric polarizability $\alpha$ and magnetic polarizability $\beta$ tell us how the nucleon responds to an external electromagnetic field,
       with current PDG values $\alpha=11.2(4)\times 10^{-4}$ fm$^3$ and $\beta=2.5(4)\times 10^{-4}$ fm$^3$ for the proton~\cite{Agashe:2014kda}.
       The polarizabilities are proportional to the volume and their smallness indicates that the proton is a rigid object due to the strong binding of its constituents.
       Whereas $\alpha+\beta$ is constrained by a sum rule, the small value for $\beta$ is commonly believed to be due to a cancellation
       between the nucleon `quark core' and the interaction with its pion cloud.

       The polarizabilities are encoded in the nucleon Compton scattering (CS) amplitude $N\gamma^\ast\to N\gamma^\ast$ which
       has many applications also beyond polarizabilities.
       The integrated CS amplitude is relevant for two-photon corrections to nucleon form factors~\cite{Arrington:2011dn}
       and perhaps also for the proton radius puzzle~\cite{Hagelstein:2015egb}.
       So far, our knowledge of the CS amplitude is restricted to a few kinematic limits including the
       (generalized) polarizabilities in real and virtual CS~\cite{Downie:2011mm}, the nucleon structure functions in the forward limit,
       and deeply virtual CS (DVCS) from where generalized parton distributions are extracted~\cite{Guidal:2013rya}.
       In addition, the crossed process $p\conjg{p}\to\gamma\gamma$ will be measured by PANDA.

       While lattice calculations for polarizabilities are underway (see~\cite{Hagelstein:2015egb} for references), the main theoretical tools to address CS are `hadronic' descriptions such as chiral perturbation theory, which provides
       a systematic expansion of the CS amplitude at low energies~\cite{Griesshammer:2012we}, and dispersion relations with a direct link to experimental data~\cite{Drechsel:2002ar,Schumacher:2013hu}.
       On the other hand, handbag dominance in DVCS is well established and a key ingredient to factorization theorems.
       Is it then possible to connect these two facets by a common, underlying approach at the level of quarks and gluons
       that reproduces all established features, from hadronic poles to the handbag picture?
       In the following we will briefly outline such an approach
       and present first calculated results for the proton polarizabilities $\alpha$ and $\beta$.

            \begin{figure*}[t]
            \centerline{%
            \includegraphics[width=0.8\textwidth]{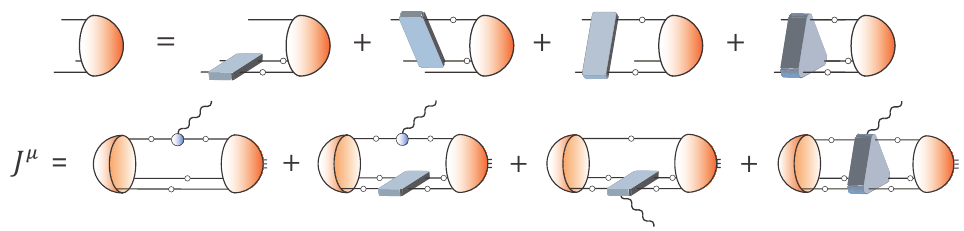}}
            \caption{Three-quark Faddeev equation (top) and electromagnetic current matrix element (bottom).}
            \label{fig:faddeev}
            \end{figure*}

 \section{The covariant Faddeev approach} \label{sec:faddeev}

             Our tool of choice is the covariant three-body Faddeev approach established in~\cite{Eichmann:2009qa}.
             Its basic equations are illustrated in Fig.~\ref{fig:faddeev}.
             The Faddeev equation determines the nucleon mass and bound-state amplitude (its `wave function') by summing up all possible
             two- and three-body interactions between dressed quarks. The electromagnetic current matrix element couples
             the photon to all microscopic ingredients and thereby satisfies electromagnetic gauge invariance.

            \begin{wrapfigure}{R}{0.35\textwidth}
            \centerline{%
            \includegraphics[scale=0.10]{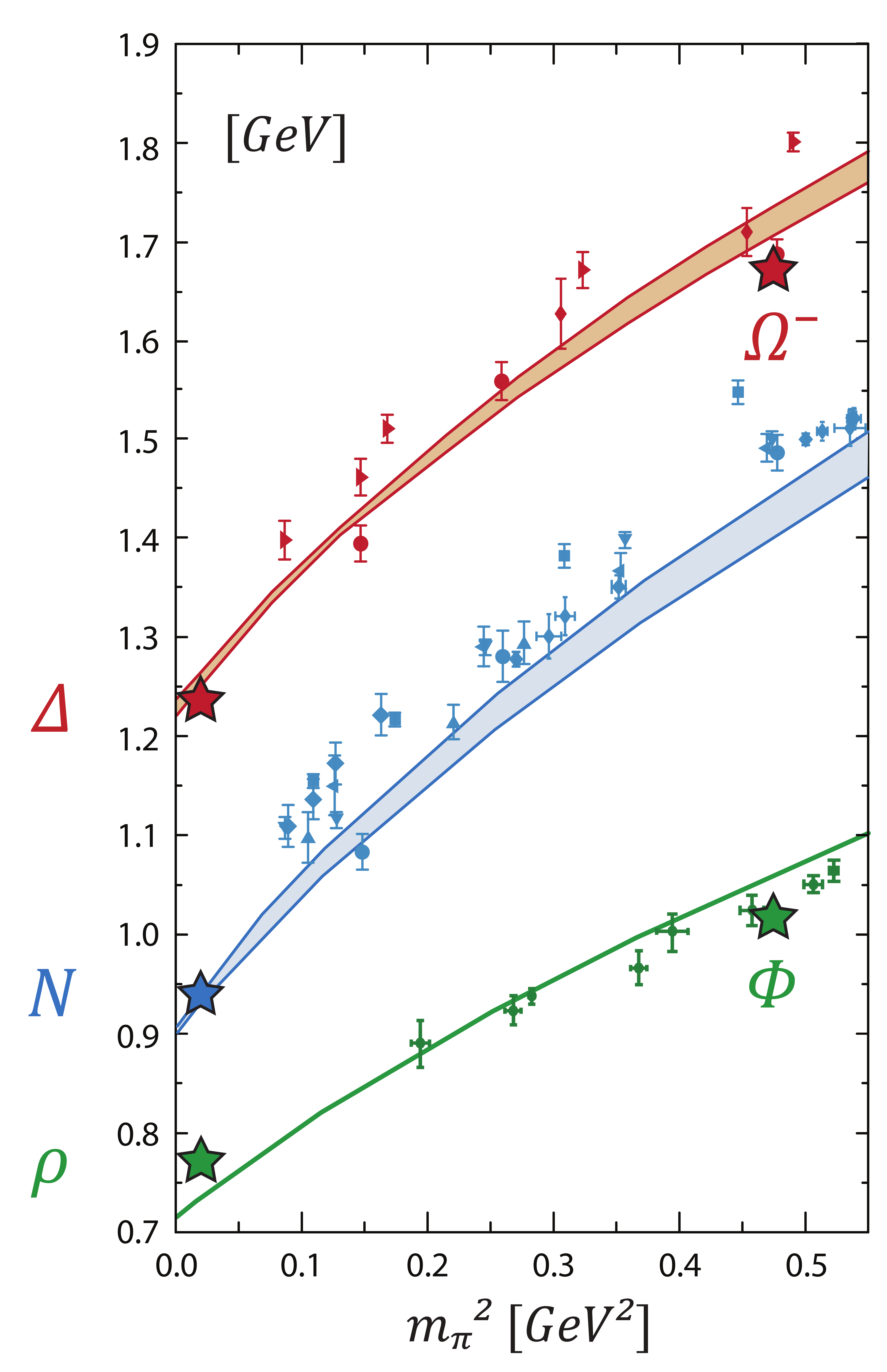}}
            \caption{$\rho-$meson~\cite{Maris:1999nt}, nucleon and $\Delta$ masses~\cite{Eichmann:2009qa}
                     calculated from their Bethe-Salpeter and Faddeev equations.
                     Stars are PDG values and symbols with error bars are lattice data (see~\cite{Eichmann:2009qa} for references).}
            \label{fig:masses}
            \end{wrapfigure}

             To solve the Faddeev equation one needs to specify its input. Three-body interactions have been neglected so far,
             and most studies have employed a rainbow-ladder truncation where the two-body kernel is given by a dressed gluon exchange.
             The dressed quark propagator is solved from its Dyson-Schwinger equation and the resulting quark mass function becomes momentum-dependent;
             it describes the transition from the input current-quark mass at large momenta to a nonperturbative, dressed `constituent quark' mass of a few hundred MeV
             in the infrared. In general, any truncation must preserve chiral symmetry to ensure a massless pion in the chiral limit via the analogous Bethe-Salpeter equation;
             see e.g.~\cite{Williams:2015cvx} for recent advances in this area.

             Whereas the applicability of rainbow-ladder in the light-meson sector is mainly limited to pseudoscalar and vector mesons,
             baryons fare much better: the approach reproduces the octet and decuplet ground state masses within $5-10\%$~\cite{Sanchis-Alepuz:2014sca}.
             Fig.~\ref{fig:masses} shows results for the $\rho-$meson, nucleon and $\Delta$ masses as functions of $m_\pi^2$ (which is also calculated)
             compared to lattice data and experiment. The only input is the quark-gluon interaction for the two-body kernel whose model dependence is given by the bands.
             In particular, once the model scale is set to reproduce the pion decay constant, there are no further parameters or approximations and all subsequent results are predictions.

             Apart from mass spectra, a range of form factors have been calculated as well within this setup. Among them are
             nucleon, $\Delta$ and hyperon electromagnetic form factors, the $N\to\Delta\gamma$ transition, and nucleon axial form factors~\cite{Alkofer:2014bya}.
             All these cases exhibit good overall agreement with experimental data (where available) and also lattice results at larger pion masses, with discrepancies at low $Q^2$ where pion-cloud effects
             become important. While the three-body Faddeev approach does not depend on explicit diquark degrees of freedom, it
             is conceptually close to the quark-diquark framework which typically yields  similar results
             and thereby establishes the presence of strong diquark correlations inside baryons~\cite{Eichmann:2008ef}.
             An advantage is that the approach is not limited to two- and three-body systems: using the very same building blocks, it has been recently also applied to
             tetraquarks and the muon g-2 problem~\cite{Eichmann:2015cra}.
             Given the body of results so far it is desirable to go a step further and ask: what can we learn about Compton scattering from such a microscopic perspective?

 \section{Compton scattering} \label{sec:cs}

 \begin{figure}
 \centering
   \includegraphics[width=1\textwidth]{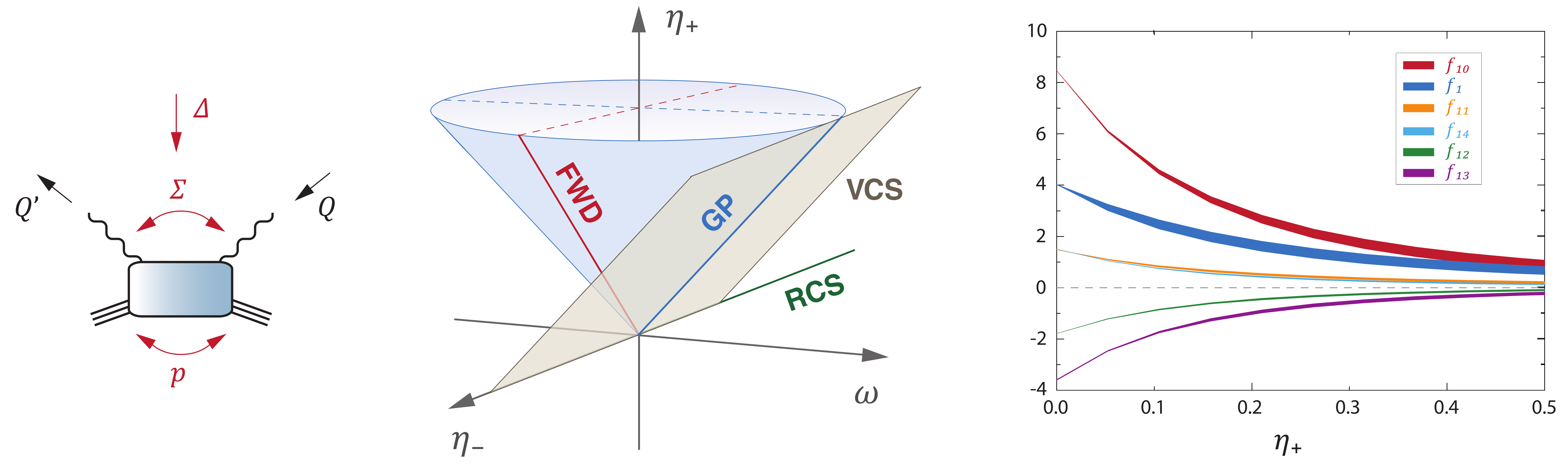}
 \caption{\textit{Left:} Kinematics and phase space in Compton scattering.
          \textit{Right:} Dominant Compton form factors corresponding to the residue of the nucleon Born terms after removing the common pole factor.
          The bands contain the full kinematic dependence on all four variables inside the cone.}
 \label{fig:phasespace}
 \end{figure}

 The nucleon CS amplitude depends on three independent momenta (see Fig.~\ref{fig:phasespace}):
 the average nucleon momentum $p$, the average photon momentum $\Sigma = (Q+Q')/2$, and the momentum transfer $\Delta=Q-Q'$.
 The process is described by four Lorentz-invariant kinematic variables which we define as
 \begin{equation}
    \eta_+ = \frac{Q^2  +  {Q'}^2}{2m^2}\,, \quad
    \eta_- = \frac{Q\cdot Q'}{m^2}\,, \quad
    \omega = \frac{Q^2-{Q'}^2}{2m^2}\,, \quad
    \lambda = \frac{p\cdot\Sigma}{m^2}\,,
 \end{equation}
 where $m$ is the nucleon mass. The kinematic phase space in the variables $\{\eta_+, \eta_-, \omega\}$ is illustrated in Fig.~\ref{fig:phasespace}.
 The spacelike region that is integrated over in nucleon-lepton scattering forms the interior of a cone around the $\eta_+$ axis.
 Its apex is where the static polarizabilities are defined, with momentum-dependent extensions to real CS ($\eta_+=\omega=0$), the doubly-virtual forward limit ($\eta_+=\eta_-$, $\omega=0$), and
 virtual CS ($\eta_+=\omega$) including the generalized polarizabilities at $\eta_-=0$.

 \vspace{-3mm}

\begin{figure}
 \centering
   \includegraphics[width=1\textwidth]{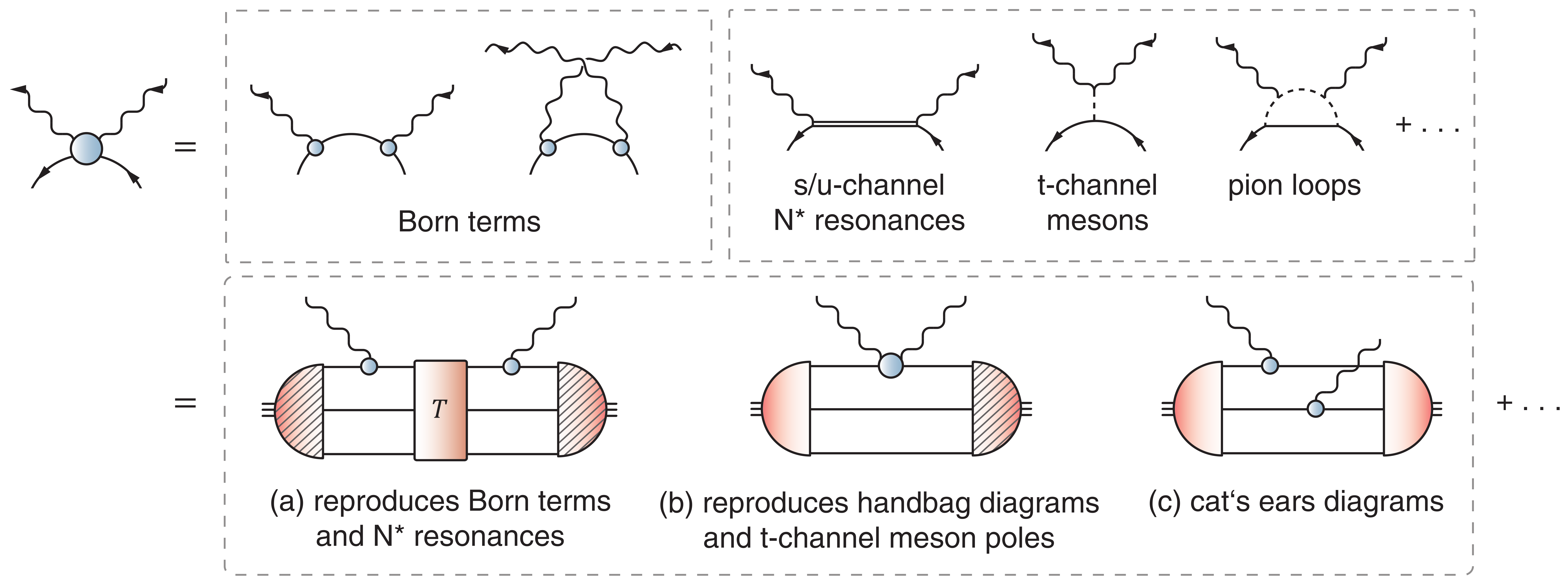}
 \caption{Hadronic vs. quark-level decomposition of the nucleon Compton scattering amplitude. The first row depicts the hadronic contributions as the sum of Born terms and a 1PI structure part.
          The latter encodes the polarizabilities and contains $s/u-$channel nucleon resonances, $t-$channel meson exchanges and pion loops.
          The second row shows the microscopic decomposition (in rainbow-ladder) featuring Faddeev amplitudes, quark propagators, quark-photon and quark Compton vertices, and the three-quark scattering matrix~\cite{Eichmann:2012mp}.}
 \label{fig:cs-decomposition}
 \end{figure}

 \paragraph{Hadronic vs. microscopic decomposition.}
 At the hadronic level the CS amplitude is given by the sum of Born terms, which are determined by the nucleon form factors,
 and a one-particle-irreducible (1PI) structure part that carries the dynamics and encodes the polarizabilities, see Fig.~\ref{fig:cs-decomposition}.
 The latter contains $s/u-$channel nucleon resonances beyond the nucleon Born terms (including the $\Delta$, Roper, etc.),
 $t-$channel meson exchanges (pion, scalar, axialvector, $\dots$), and pion loops,
 with well-established low-energy expansions in chiral effective field theory.
 This is usually viewed as the `correct' description at low energies, whereas the handbag picture is interpreted as the `correct' one at large photon virtualities.
 Hence again the question: is there a common underlying description at the quark level that is valid in \textit{all} kinematic regions and encompasses both approaches?

 In analogy to the form factor diagrams in Fig.~\ref{fig:faddeev} one can derive a closed expression for the CS~amplitude at the quark level~\cite{Eichmann:2011ec,Eichmann:2012mp}.
 The topologies that survive in a rainbow-ladder truncation (apart from permutations and symmetrizations) are collected in the second row of Fig.~\ref{fig:cs-decomposition}.
 Ambiguities stemming from intermediate offshell hadrons never arise here because hadronic degrees of freedom do not appear explicitly.
 Instead, the diagrams reproduce the onshell hadron pole contributions:
 \begin{itemize}
 \item Diagram (a) depends on the three-quark scattering matrix which contains all possible baryon poles, so it reproduces
 the nucleon Born terms as well as all $s/u-$channel resonances. \\[-3mm]
 \item Diagram (b) contains the quark two-photon (quark Compton) vertex,
 which has an analogous decomposition into quark Born terms and a 1PI part.
 The Born terms provide the handbag contributions.
 The 1PI part features a quark-antiquark scattering matrix
 that contains all possible $t-$channel meson poles and thereby reproduces the meson exchanges in the first row.
 \end{itemize}
 Neither the handbag nor the cat's-ears contributions from diagram (c) have a direct analogue in the hadronic expansion where they are rather absorbed into counterterms.
 Vice versa, the diagrams in the bottom do not contain the microscopic representation of pion loops because those only enter beyond rainbow-ladder.
 In any case, the sum of all graphs in the box satisfies electromagnetic gauge invariance  so that the resulting CS amplitude is purely transverse; it is $s/u-$channel crossing symmetric;
 it reproduces all known hadronic poles; and it contains the handbag contributions which are perturbatively (and presumably also nonperturbatively) important.

 Is it feasible to calculate all these microscopic diagrams in analogy to what has been achieved for form factors?
 The main obstacle is diagram (a): while there has been progress in the calculation of three- and four-point functions~\cite{Eichmann:2015nra},
 the treatment of six-point functions is so far beyond reach. We therefore approximate this graph by summing up the leading hadronic diagrams in the form of
 nucleon resonances. Neglecting also diagram (c), we calculate graph (b) in rainbow-ladder but without further approximations:
 the quark propagator is obtained from its Dyson-Schwinger equation, the nucleon amplitude from the covariant Faddeev equation, and the quark Compton vertex
 including all (128) tensor structures from its inhomogeneous Bethe-Salpeter equation~\cite{Eichmann:2012mp}.

 \vspace{-3mm}

 \paragraph{Gauge invariance.}
 The problem with this strategy is that only the sum of diagrams (a-c) is gauge invariant but not the individual graphs alone.
 Since transversality is connected with analyticity, a simple transverse projection does not suffice because
 an approximation that breaks electromagnetic gauge invariance can induce kinematic singularities that render its results meaningless.
 The problem can be illustrated with a textbook example, namely the photon vacuum polarization whose general form is
 $\Pi^{\mu\nu}(Q) = a\,\delta^{\mu\nu} + b\,Q^\mu Q^\nu$. The coefficients $a$ and $b$ are functions of $Q^2$ and
 must be analytic at $Q^2=0$; poles would correspond to intermediate massless particles but since the vacuum polarization is 1PI
 intermediate propagators are excluded by definition. Gauge invariance entails transversality, $Q^\mu  \Pi^{\mu\nu}=0$, which fixes $a=-b\,Q^2$.
 The vacuum polarization can then be written as the sum of a transverse part and a `gauge part' (which is \textit{not} longitudinal):
 \begin{equation}\label{vac-pol}
    \Pi^{\mu\nu}(Q) = \Pi(Q^2)\,t_{QQ}^{\mu\nu} + \widetilde{\Pi}(Q^2)\,\delta^{\mu\nu}\,, \qquad t^{\mu\nu}_{AB} = A\cdot B\,\delta^{\mu\nu} - B^\mu A^\nu\,.
 \end{equation}
 The transverse dressing function $\Pi(Q^2)$ is free of kinematic singularities and zeros.
 The gauge part $\delta^{\mu\nu}$ is the tensor that we eliminated in the first place, so $\widetilde{\Pi}(Q^2)$ must vanish due to gauge invariance.
 This is what happens in dimensional regularization, whereas a cutoff breaks gauge invariance and induces a quadratic divergence (only) in the gauge part.
 If we did not know about the decomposition~\eqref{vac-pol} and performed a transverse projection, $\Pi(Q^2)$ would pick up a $1/Q^2$ pole from the gauge part
 which invalidates the extraction of $\Pi(Q^2=0)$.
 The transverse/gauge separation is also convenient if gauge invariance is broken by more than a cutoff, for instance by an incomplete calculation:
 ultimately the sum of all gauge parts must vanish, but the partial result for $\Pi(Q^2)$
 is still free of kinematic problems and --- ideally --- not strongly affected by gauge artifacts.

 This simple example also provides the template for the CS amplitude, where a complete decomposition into transverse and gauge parts free of kinematic singularities and zeros is necessary as well:
 \begin{equation}\label{cs-basis}
   \mathcal{M}^{\mu\nu}(p,Q',Q) = \frac{1}{m}\,\conjg{u}(p_f)\, \bigg[ \, \underbrace{\left( \frac{f_1}{m^4}\,t^{\mu\alpha}_{Q'p}\,t^{\alpha\nu}_{pQ} + \frac{f_2}{m^2}\,t^{\mu\nu}_{Q'Q} + \dots\right)}_{\text{transverse part, 18 tensors}}
                                                                   \;\; +  \underbrace{\left(\dots\right)}_{\text{\scriptsize gauge part,} \atop \text{\scriptsize 14 tensors}} \,\bigg] \,u(p_i)\,,
 \end{equation}
 with $t^{\mu\nu}_{AB}$ defined in Eq.~\eqref{vac-pol}.
 We employ the transverse tensor basis of Refs.~\cite{Tarrach:1975tu,Drechsel:2002ar} but
 insert factors of $\omega$, $\lambda$ and $m$ where necessary, so that all Compton form factors (CFFs) $f_i(\eta_+, \eta_-, \omega, \lambda)$ are
 dimensionless and invariant under photon crossing and charge conjugation.
 The gauge part must be zero and vanish if all diagrams in Fig.~\ref{fig:cs-decomposition} are included.
 However, even if one breaks gauge invariance by retaining only a subset of diagrams, the transverse CFFs
 still yield a well-defined prediction.

 This can be understood already at the hadronic level. The definition of polarizabilities
 entails that both the Born and 1PI parts must be individually gauge invariant,
 so the expansion~\eqref{cs-basis} must hold for both contributions alone.
 The Born terms are specified by the nucleon's electromagnetic current,
 but since the intermediate nucleon is offshell the half-offshell nucleon-photon vertex can have more tensor structures.
 It is well known that only an onshell Dirac current with $Q^2$-dependent Pauli and Dirac form factors ensures gauge invariance
 of the Born term.\footnote{For arbitrary offshell form factors the Born terms must be combined with (non-diagrammatic) parts
 of the 1PI contribution to arrive again at a gauge-invariant expression, see~\cite{Eichmann:2012mp} for a discussion.}
 The right panel in Fig.~\ref{fig:phasespace} shows the leading CFFs for this case after removing the common nucleon pole factor.
 Note that all CFFs are well-behaved and approach constant values for $\eta_+\to 0$. As required, the gauge part is exactly zero.
 The implementation of offshell form factors destroys this property: the gauge part then no longer vanishes but within a reasonable range of model parametrizations the
 transverse CFFs remain almost unchanged.

 Another remarkable feature is visible in Fig.~\ref{fig:phasespace}: the bands contain the \textit{full} kinematic dependence
 on all four variables $\eta_+$, $\eta_-$, $\omega$ and $\lambda$ inside the cone, but effectively they only depend  on $\eta_+$.
 The residues of the nucleon Born terms therefore scale with $\eta_+$, which reflects the symmetric makeup of the phase space.
 The hadronic poles form planes in the phase space that will generally counteract this symmetry property:
 the nucleon Born poles appear at $\eta_-=\lambda=0$;
 the nucleon resonance poles form vertical planes at fixed $\eta_-<0$, where the value of $\lambda$ depends on the width of the resonance;
 and $t-$channel meson poles appear at fixed $\Delta^2 =-m_i^2$ $\Rightarrow$ $\eta_-=\eta_++m_i^2/(2m^2)$.

 \begin{figure}
 \centering
   \includegraphics[width=0.8\textwidth]{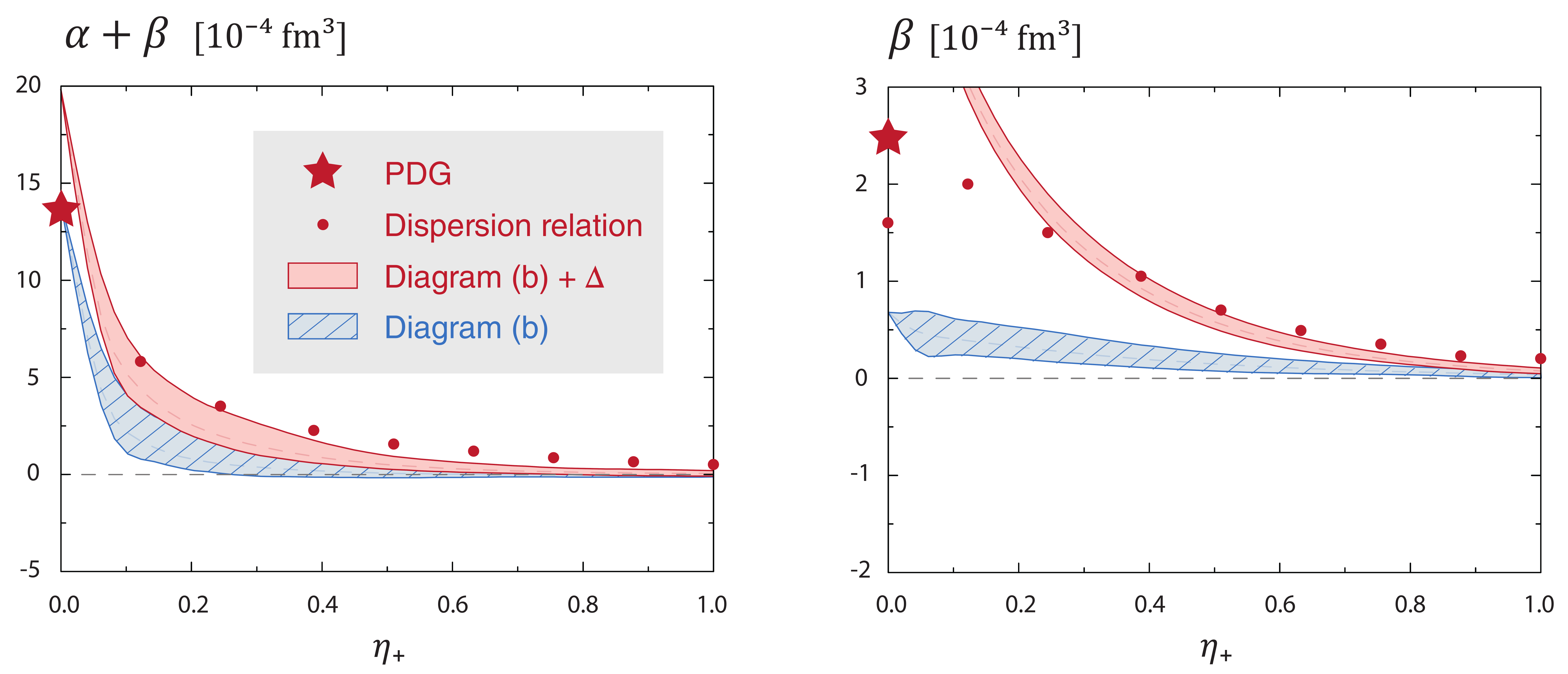}
 \caption{Proton polarizabilities as functions of $\eta_+$. The bands were obtained from diagram (b) with and without the $\Delta$ contribution.
          The dots were extracted from Ref.~\cite{Downie:2011mm} and the stars are the experimental values~\cite{Agashe:2014kda}.}
 \label{fig:polarizabilities}
 \end{figure}

 \vspace{-3mm}

 \paragraph{Polarizabilities.}
The nucleon polarizabilities $\alpha$ and $\beta$ are related to the CFFs $f_1$ and $f_2$ in the limit where all kinematic variables are zero:
 $\{\alpha+\beta,\beta\} = \{f_1, f_2\} \times \alpha_\text{QED}/m^3$. In Fig.~\ref{fig:polarizabilities} we show preliminary results from the quark-level calculation
 extracted from the basis in Eq.~\ref{cs-basis}. So far they are only ballpark estimates:
 the quark Compton vertex that enters in the calculation depends on 6 Lorentz invariants and 128 tensor structures
 and its transverse/gauge separation is extremely sensitive to the numerics.
 We extracted the momentum dependence of $\alpha+\beta$ from $f_1$ only whereas the standard definition~\cite{Drechsel:2002ar} contains admixtures from higher CFFs at $\eta_+>0$,
 but for those our results are still too noisy.

 The hatched bands in Fig.~\ref{fig:polarizabilities}  are the outcome of diagram (b) inside the cone. For
 the total result we added the $\Delta$ resonance, the dominant hadronic contribution to diagram (a),
 using a parametrization for the experimental $N\to\Delta\gamma$ form factors. 
 For comparison we  plot the dispersion relation results for the generalized polarizabilities from Refs.~\cite{Downie:2011mm,Drechsel:2002ar}.
 The figure makes clear that the sum $\alpha+\beta$ is dominated by diagram (b) and, as it turns out, especially by the handbag contributions.
 The magnetic polarizability $\beta$ is dominated by the $\Delta$ pole from diagram (a) whereas (b) contributes little due to cancellations.
 The discrepancy at low $\eta_+$ is presumably due to missing pion loops --- $\beta$ is subject to cancellations between the quark core
 (which then mainly comes from the $\Delta$ pole) and pion cloud effects.

 To summarize, we demonstrated how to extract microscopic information on nucleon polarizabilities from the decomposition in Fig.~\ref{fig:cs-decomposition}.
 It will be further interesting to investigate spin polarizabilities and gather knowledge on the spacelike momentum dependence of the CS amplitude,
 which will improve our understanding of two-photon corrections to form factors as well as the proton radius puzzle.
 Finally, the same framework can be adapted to other processes such as pion electroproduction,
 which has contributed much to our knowledge of nucleon resonances and transition form factors.
 For their clean extraction one needs to know the non-resonant `QCD background' beyond hadronic exchanges, which
 is information that a microscopic approach can provide.

 \vspace{-3mm}

  \begin{acknowledgements}
  I am grateful to C. S. Fischer and G. Ramalho for valuable discussions, and I would like to thank the organizers of the \textit{Light Cone 2015} conference for their support.
  \end{acknowledgements}

 \bibliographystyle{spbasic}


 \end{document}